\documentclass[oneside]{book}
\usepackage[english]{babel}
\usepackage{amsfonts}
\usepackage{graphicx}

\usepackage{caption}
\usepackage{xcolor}

\usepackage[a4paper,right=25mm,left=25mm,top=25mm,bottom=25mm,headsep=10mm]{geometry}
\usepackage{fancyhdr,lipsum}
\usepackage{amssymb}
\usepackage{subfig}
\usepackage{algorithm}
\usepackage{algpseudocode}

\usepackage{float}
\usepackage{longtable}

\usepackage[T1]{fontenc}

\usepackage[hidelinks]{hyperref}
\usepackage[document]{ragged2e}

\usepackage{arabtex} 
\usepackage{utf8}

\usepackage{pifont}  % for some special characters 

\usepackage[nonumberlist,acronym]{glossaries}
\makeglossaries

\setcode{utf8}

\addto{\captionsenglish}{}

\begin{document}

\begin{Large}
\pagenumbering{roman}
\justifying

%=========================== Cover Page :=====================
\thispagestyle{plain}
\thispagestyle{empty}
\begin{figure}
    \centering
    \includegraphics[scale=0.5]{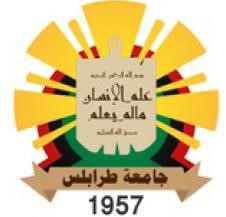}
\end{figure}

\begin{center}
{\Large \rm University of Tripoli \linebreak}
{\Large \rm Faculty of Engineering \linebreak}
{\Large \rm Electrical and Electronic Engineering Department \linebreak}
\vspace*{1cm}
\baselineskip 30pt
\end{center}

\vspace{-0.2cm}
\begin{center}
    {\Large \rm B.Sc. Project \linebreak}
\end{center}
\vspace{-0.2cm}

\begin{center}
{\huge \rm \textbf{Smart 6G Sky for Green Mobile IOT Networks}}
%Multi-Objective Optimization for UAV-Assisted Energy Harvesting powered IoT Networks Based on DDPG Algorithm}}
\baselineskip 30pt
\end{center}
\begin{center}
\baselineskip 30pt
\center{\Large \rm Prepared by: Qusai Fathi Bshioah}\vspace*{-.5cm} \center{\Large \rm Supervised by: Dr. Nadia Adem}
\end{center}
\begin{center}
\baselineskip 30pt
{\large \rm Fall 2021 \linebreak}
{\large \rm Tripoli-Libya \linebreak}
\end{center}
% ===================== end cover page ===============
% ====================  start Acknowledgment ========
\newpage
\thispagestyle{plain}
\begin{center}
    \Large {\bf \uppercase{Acknowledgements}}
\end{center}
\justifying
\vspace{3\baselineskip}

I would like to thank my supervisor Dr. Nadia Adem, for the guidance, encouragement and advice she has provided as her student, caring so much about my work and responding to my questions and queries so promptly.\\
Words cant express how I am extremely thankful to my parents for their unconditional love, endless prayers, caring and immense sacrifices for educating and preparing me for my future.\\
I am greatly indebted to my honorable teachers of the  Department of Electrical and Electronic Engineering at the University of Tripoli who taught me during the course of my study.  Without any doubt, their teaching and guidance have completely transformed me to the person that I am today.\\
I would like to say thanks to my friends  for their kind support and care.\\
Finally, I would like to thank all the people who have supported me to complete
the project work directly or indirectly.

\newpage

%=========================== End Acknowledgment===========
%======================= Start Abstract page ============
\thispagestyle{plain}
\begin{center}
    \huge{\textbf{Abstract}}
\end{center}
\justifying
\vspace{1\baselineskip}

6G is envisioned to connect everything and yet to be a hundred times more energy efficient than the 5G.  Thanks for its ability to use sources of ambient energy, energy harvesting (EH) is promising in alleviating the challenge of meeting such conflicting demands.  Moreover, when it comes to the Internet of things (IoT),   one of the foundations for enabling connecting everything, the need for EH may become inevitable.  IoT involves connecting not only  devices that are large in number, but also hard to reach.  The good news, nevertheless, is that the unmanned aerial vehicle (UAV), owning to its flexibility and ease of deployment is emerging to offer   communication services when infrastructure is lacking.  Merging the UAV and IoT is of quite interest as the former could not just enable flexible connectivity for the IoT but also  powering them in spite of any restrictions. However, managing the UAV assisted IoT resources to meet certain data communications and EH quality measures while keeping the UAV consumed energy minimized is a major challenge as this corresponds to a non-convex optimization problem. Things, obviously, become even worse when the IoT network devices are mobile. Owing to the success of artificial intelligence (AI) in solving  complicated problems, in this project we rely on  the deep deterministic policy gradient (DDPG) technique, to manage the UAV assisted IoT resources.  Our results show that   DDPG  achieves joint optimization of three objectives, namely sum data rate and harvested energy maximization, and energy consumption minimization, while out performing traditional mathematical schemes. The code of this project is made publicly accessible at \href{https://github.com/QusaiBshiwa/Smart-6G-Sky-for-Green-Mobile-IOT-Networks}{https://github.com/QusaiBshiwa/Smart-6G-Sky-for-Green-Mobile-IOT-Networks}.%{Smart 6G Sky for Green Mobile IOT Networks}.}

\thispagestyle{plain}
\begin{center}

    \vspace{5cm}
    \justifying
    \begin{RLtext}
    
    % \begin{\huge}
    \center{\huge{\textbf{ملخص}}}
    \vspace{1\baselineskip}
    الرؤية المستقبلية للجيل السادس من انظمة الاتصالات اللاسلكية هي ربط كل شئ بكفاءة استخدام للطاقة اكبر بمئة مرة من الجيل الخامس. بفضل مقدرتها علي استخدام مصادر الطاقة المتوفرة في البيئة المحيطة, تعد تقنية حصاد الطاقة حلاً واعدًا للتحدي المتمثل في تلبية المتطلبات المتداخله لهذه الانظمة. علاوة على ذلك, عندما يتعلق الامر بتقنية انترنت الأشياء, والتي تعد احدى أسس ربط كل شئ, فقد  تصبح الحاجة لتقنية حصاد الطاقة امرا لا مفر منه.  انترنت الأشياء لا يتضمن فقط ربط عدد كبير من الأجهزة, ولكن ربط الأجهزة التي يصعب الوصول اليها ايضا. لحسن الحظ, أن المركبة الجوية غير المأهولة, نتيجة لمرونة استعمالها و سهولة اطلاقهاو بدأت في الظهور كوسيلة لتقديم خدمات الجيل السادس. ربط الطائرات بدون طيار وإنترنت الأشياء ذا أهمية كبيرة لأن الأول لا يمكنه فقط تمكين الاتصال المرن  لهذه الاجهزة ولكن أيضا تزويدها بالطاقة  لتشغيلها بغض النظر عن القيود التى قد تكون لديهم في التوزيع. ومع ذلك ، فإن إدارة موارد انترنت الأشياء المدعومة من الطائرات بدون طيار لتحسين اتصال البيانات و جودة الطاقة المحصودة مع المحافظة على الطاقة المستهلكة للطائرة بدون طيار إلى أدنى حد يمثل تحديًا كبيرًا لكون كل منهم جزءا متعارضا مع الاخر ممَََ يجَعلها مشكلة معقدة. أيضا الامور تزداد سوءا عندما تكون أجهزة شبكة إنترنت الأشياء متحركة. نظرًا لنجاح الذكاء الاصطناعي في حل مثل هذه المشكلات المعقدة ، نعتمد في هذا المشروع على التطورات الحديثة في أساليب الذكاء الاصطناعي ، وهي تقنية التدرج الحتمي العميق للسياسة ، لإدارة مشكلة إدارة موارد إنترنت الأشياء بمساعدة الطائرات بدون طيار. تظهر نتائجنا أن هذه التقنية تحقق تحسينًا مشتركًا لثلاثة أهداف ، وهي مجموع معدل البيانات وزيادة مقدار الطاقة المحصودة ، وتقليل استهلاك الطاقة للمركبة الجوية غير المأهولة , مع مقدرة التفوق على طرق الرياضيات التقليدية
    \end{RLtext}
    \begin{RLtext}

    \end{RLtext}
\end{center}
\justifying
% \vspace{3\baselineskip}

%==========================End Abstract page =========
%==========================Start table of content =========
\tableofcontents
%==========================End Table of content =========
%========================== Start List of figures======
\listoffigures
\listoftables

%================== End list of figures==============
%====================== start Acronyms ================
% \begin{abbreviations}
\newacronym{iott}{IoT}{ Internet of things}
\newacronym{xr}{XR}{Extended Reality}
\newacronym{uav}{UAV}{Unmanned Aerial Vehicle}
\newacronym{bs}{BS}{Base Stations}
\newacronym{eh}{EH}{Energy harvesting}
\newacronym{ai}{AI}{Artificial Intelligence}
\newacronym{dc}{DC}{Data Collection}
\newacronym{ml}{ML}{Machine Learning}
\newacronym{nn}{NN}{Neural Networks }
\newacronym{dl}{DL}{Deep Learning }
\newacronym{drl}{DRL}{Deep Reinforcement Learning }
\newacronym{rl}{RL}{Reinforcement Learning }
\newacronym{dqn}{DQN}{Deep Q-Network }
\newacronym{ddpg}{DDPG}{Deep Deterministic Policy Gradient}
\newacronym{moddpg}{MODDPG}{Multi-Objective Deep Deterministic Policy Gradient}
\newacronym{mdp}{MDP}{Markov Decision Processes}
\newacronym{moo}{MOO}{Multi-Objective Optimization}

\newacronym{los}{LoS}{Line of Sight}
\newacronym{nlos}{NLoS}{Non-Line of Sight}

\newacronym{rf}{RF}{Radio Frequency}
\newacronym{qoss}{QoS}{Quality of Service}
\newacronym{relU}{ReLU}{Rectified Linear Unit}

\glsaddall
\printglossary[type=\acronymtype]

\printglossary

%======================== End Acronyms===============

%============================== Chapter 1 =====================================

\chapter{Introduction}
\pagenumbering{arabic}

Internet of things (IOT), with its ability to facilitate the interconnectivity of physical world by enabling a global network of devices to communicate and interact with users, has been going beyond entertaining people and adding convenience to their life by connecting different home objects for example, to improving economy as it has been playing significant part in some societies  manufacturing, logistics, etc.    Furthermore this technology has a variety of military and national securities related applications including borders protection.  When it comes to medical and healthcare, in addition, IOT technology can be considered a life saver.  In spite of the role the IOT may play in every element of modern life, the wirelessly connected IOT devices   are in most applications large in number and physically hard to reach hence powering them holds back this technology from achieving its full potential.  Whilst providing a solid uninterrupted source of power, battery technology suffers from a range of drawbacks when used as part of IoT devices. These drawbacks can include the requirement of frequently recharging devices, potential pollution of the environment from damaged cells and degradation of the battery technology itself over a large number of charge cycles.\\
The use of renewable energy sources captured from outside sources through energy harvesting (EH) technology is an attractive alternative to battery technology as it often mitigates most of these drawbacks, but often cannot be used as a swap-in solution to replace conventional battery-based systems. This is because energy harvesting supplies are very dependent on the environment in which they are situated and need to be managed intelligently alongside the load they are powering to allow for reliable operation ~\cite{Prauzek}.

%+++++Motivation section++++++++++++++++++++++++++++++++
\section{Motivation}
The fifth generation (5G) of wireless systems is marketed as a provider of IoT services, but the ongoing deployment of 5G has resulted in it being unable to meet the technical standards of IoT networks and their rapidly expanding services. The shortcomings of 5G are driving a global effort to define the next-generation 6G wireless communication system, a system that is truly capable of delivering these far-reaching applications while employing new technologies to address these challenges ~\cite{WalidSaad}. One of these challenges is the requirement for long-lasting and compact batteries, which can limit the use of such devices. Therefore, energy harvesting could be a solution for making IoT devices autonomous, allowing for widespread use of these systems in a variety of applications ~\cite{MuhammadImran}.  However, relying on the current terrestrial cellular system to achieve such a goal will be at a much more difficulty level.  Current cellular networks are overly engineered and any investments in improving them will undoubtedly be exorbitant.  Unmanned Aerial Vehicle (UAV) networks, with their flexible mobility, ease of deployment, and low cost offer to extend, boost, or replace terrestrial Base Stations (BS) by providing the opportunity to adaptively yet fairly and efficiently manage resources in real time to meet the dynamic-heterogeneous-massive needs.  However, traditional methods are incapable of dealing with such complex networks.\\
Fortunately, Artificial Intelligence (AI) technologies have a high potential for dealing with multi-state network statuses and demands.  AI techniques have gained attraction in the wireless networking community after demonstrating their effectiveness in solving problems with a high degree of freedom in a variety of fields.
Reinforcement Learning (RL) methods, for example, in which decisions (or actions) that satisfy certain criteria can be learned in a given environment without any prior knowledge.  However, RL has been used in a variety of applications, including security and spectrum management. ~\cite{ling,ortiz} and they also fail to optimize a solution as the state space that represents the problem grows in size. This can be easily solved by combining Neural Networks (NN) and Reinforcement Learning (RL), resulting in Deep Reinforcement Learning (DRL), an emerge that has the potential to optimize the prediction of actions for any number of states, providing an enormous range of potentials for solving such non-convex problems efficiently.

\section{Related Work}
The existing literature has studied a number of problems focused on \acrshort{eh} communication systems when non-causal knowledge of the \acrshort{eh} process is assumed such as \cite{Yener}. Their approach assumes that the amount of energy and the arrival time of this energy is known at the beginning of each communication . Although this assumption allows the ease of calculations but it can't be fulfilled in reality.\\
More realistic approaches assume statistical information about \acrshort{eh} process. In \cite{Ozel} the authors investigated a point to point communication in \acrshort{eh} nodes scenario with the assumption of a fading channel, their  approach of solving this problem is using continuous time random dynamic programming with statistical and causal knowledge of the energy and fading variations.\\
All the aforementioned approaches require knowledge of the statistics of the \acrshort{eh} process, which in practical scenarios might not be available.  Consider for example, a stationary \acrshort{eh} transmitter that transfers energy to different devices simultaneously.  In this scenario, the \acrshort{eh} process cannot be considered as stationary and consequently, keeping track of its statistics becomes challenging.  Moreover, these stationary base stations might be difficult to be carried out in these networks without them having an excessive cost and them being over-engineered.  Therefore, \acrshort{uav} assisted communication could provide solutions for these networks by exploiting the several advantages of \acrshort{uav}s such as the simplicity of deployment, lower cost, high-altitude assisted transmission and so on \cite{gupta}.\\
However, the \acrshort{uav} is a battery powered communication terminal, Therefore, energy saving is as an important metric in designing future wireless communication systems.  The authors in \cite{li} have studied the energy minimization of the \acrshort{uav} in various communication systems, However, their work only focus on minimizing the communication energy consumption as in the conventional terrestrial wireless communication, for \acrshort{uav} in practice it was found that the communication energy consumption is usually much lower compared to propulsion energy consumption, which is required to maintain the \acrshort{uav}  aloft and enable their mobility. Therefore this propulsion energy consumption and trajectory design became a dominant factors that needs to be taken in consideration for achieving energy-efficient UAV communications.   To this end, the authors in \cite{zeng} developed a mathematical model for the propulsion energy consumption of the \acrshort{uav}.  From IoT applications perspective, timely data collection is crucial to the accuracy and reliability of derived decisions~\cite{kaul}. The authors in \cite{Zhang2017} investigated the energy consumption and data collection rate trade-off between the UAV and its served devices via UAV trajectory design revealing that the the closer the UAV flies to each device, the less energy is needed to transmit its data. However. 
Merging the mobility of \acrshort{uav}, the randomness and dynamics of  \acrshort{iott} system and energy harvesting process pose great challenges to the optimization of \acrshort{uav}-assisted wireless \acrshort{iott} networks.  Facing the complex and dynamic \acrshort{iott} network environments, The \acrshort{uav} is required to be equipped with the ability to sense its surrounding and the ability of taking real-time decision, therefore, traditional optimization methods rapidly become unmanageable for these sophisticated network optimizations. Recently, \acrfull{ai} has been considered as the major innovative technique for \acrshort{uav}-assisted \acrshort{iott} system \cite{zhang}.  Specifically, \acrfull{drl} has become a promising technology and has attracted extensive attention. Taking full advantage of \acrshort{drl} algorithm, \acrshort{uav} can learn to build knowledge about the massive \acrshort{iott} environment without knowing the complete network information through iterative interaction, and then modifies its action strategy accordingly, one of the most trending algorithms of \acrshort{drl} is \acrshort{ddpg} which was was leveraged for the continuous control tasks~\cite{hunt}. The authors in \cite{yang} suggested a DDPG technique for learning UAV control policies with multiple goals. 

\section{Contributions}
In this project, we aim, through the use of the most recent advances in AI, to jointly optimize UAV and IOT energy management while meeting some other communication quality of service requirements. 
In our model, UAV is allowed to allocate in 2-D. The devices can be line-of-sight (LoS) or even none-line-of-sight (NLoS). Moreover, unlike previous studies mentioned above that consider the deployment of the UAVs assuming static users, the AI algorithm will be trained to deploy UAVs with dynamic  users .The main contributions of this project are summarized as follows.
 
\begin{enumerate}
 \item We implement and apply the DDPG technique to efficiently and jointly optimize the total data rate and harvested energy while minimizing the UAV's energy consumption for mobile IoT devices.  This project is  built on and extends the work of~\cite{Yu2021} in which a DDPG based framework is provided to solve for energy management of  static IoT devices. %Which is different from
 
 \item We demonstrate that the AI algorithm is more effective than traditional mathematical techniques %e.g. the one provided in~\cite{Zhang2017} 
 in solving our defined non-convex problem, in spite of the fact that existing baselines (e.g. the one provided in~\cite{Zhang2017})  consider a much more simplified system than ours including the presence of LoS links with users and their stationarity.

\end{enumerate}

%++++++++++++++start of organization ++++++
\section{Organization}
The remainder of this project is structured as follows: Chapter 2 introduces the fundamental concept of energy harvesting and its applications. The network modeling and problem presentation in Chapter 3 we discuss DDPG in detail in Chapter 4. In Chapter 5, we show and analyze simulation results, and in chapter 6, we present the project  conclusion.
%+++++++++++++ End of organization+++++++++++++

%=============================End of chapter 1=================================

%==========================Start of chapter2=================================
\chapter{Green Communications}
\section{The Concept of Energy Harvesting}
Because of advancements in wireless networks, some applications require devices to have a long lifetime. Traditional batteries are not always advantageous because they require human intervention to be replaced. As a result, obtaining the electrical power required to operate these devices is a major concern. Therefore. Alternative energy sources to traditional batteries must be considered. The electrical energy required to power these devices can be obtained by harvesting the energies present in the surrounding environment. This process contributes to the provision of unlimited energy for the duration of the electronic device's lifespan.  Therefore, the process of extracting energy from the ambient environment and converting it into consumable electrical energy is known as energy harvesting,\\
The energy harvesting sources can be used to increase the lifetime and capability of the devices by either replacing or augmenting the battery usage. The devices powered by energy harvesters can be used to provide vital information on operational and structural circumstances by placing them in inaccessible locations.
\section{Sources of Energy Harvesting}
The classification of energy harvesting can be organized on the basis of the form of energy they use to harvest the power. The various sources for energy harvesting
are wind turbines, photovoltaic cells, thermoelectric 
generators and mechanical vibration devices (piezoelectric) ~\cite{Farrar}. Table \ref{harvesters} shows some of the harvesting methods with their power
generation capability
\begin{table}[H] \label{harvests}
    \centering

        \caption{ENERGY HARVESTING SOURCES}
       
   \resizebox{0.55\textwidth}{!}{ \begin{tabular}{|c|c|}
  
    \hline
         \textbf{Harvesting Method}  & \textbf{Power Density}   \\\hline 
           Solar Cells & 15mw/$cm^{3}$ \\\hline
           Piezoelectric & 330$\mu W /cm^{3}$ \\\hline
           Vibration & 116 $\mu W /cm^{3}$ \\\hline

       \hline 
    \end{tabular}}

        \label{harvesters}
\end{table}
The general properties to be considered to characterize a portable energy supplier are described by \cite{Fry}. The list includes electrical properties such as power density, maximum voltage and current; physical properties such as the size, shape and weight; environmental properties such as water resistance and operating temperature range. 

\section{Methods of Harvesting}
\subsection{Mechanical Vibration}
When a device is subjected to vibration, an inertial mass can be used to create movement. This movement can be converted to electrical energy using three mechanisms: piezoelectric, electrostatic and electromagnetic. The form of energy utilized here is the mechanical energy.

\begin{figure}
    \centering
    \includegraphics[scale = 0.7]{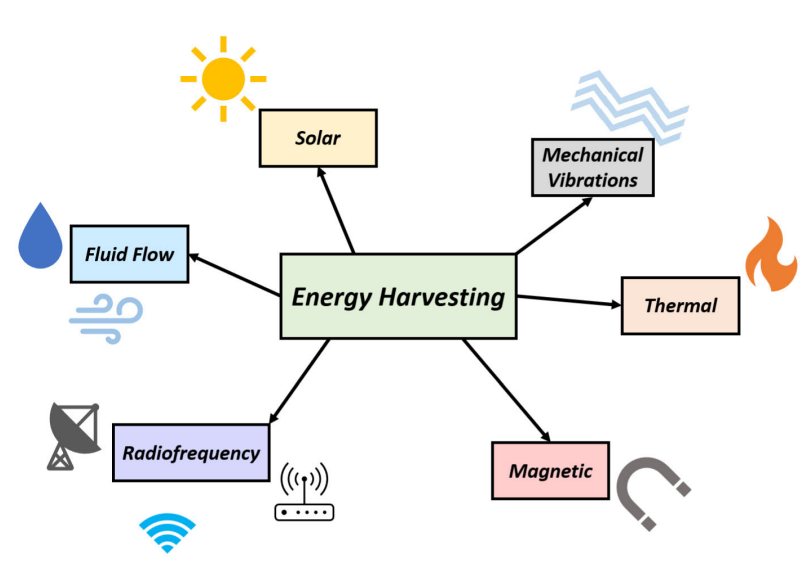}
   
        \caption{Sources of energy that can be harvested}

    \label{ehmeth}
\end{figure}

\subsubsection{\large Piezoelectric Materials }
The materials that convert mechanical energy from pressure, vibrations or force into electricity are called Piezoelectric Materials. They are capable of generating electrical charge when a mechanical load is applied on them. This property of piezoelectric materials is considered by the researchers to develop various piezoelectric harvesters in order to power different applications. Due to their inherent ability to detect vibrations, piezoelectric materials have become a viable energy harvesting sources. Some are naturally occurring materials such as quartz.  Polycrystalline ceramic is a common piezoelectric material. since they shows a high efficiency of mechanical to electrical energy conversion.\\
Using piezoelectric materials to harvest energy requires a mode of storing the energy generated. This means they can either implement a circuit used to store the energy harvested or a circuit developed to utilize the energy harvested in producing excess energy then the energy harvested can be stored in rechargeable batteries instead of using capacitors to store the energy ~\cite{Sodano}.\\
The properties of piezoelectric materials vary with age, stress and temperature. The possible advantages of using piezoelectric materials are the direct generation of desired voltage since they do not need a separate voltage source and additional components. Some disadvantages are that piezoelectric materials are brittle in nature and sometimes allow the leakage of charge.

\subsubsection{\large Electromagnetic Energy Harvesting }
Electromagnetic energy harvesting can be achieved by the principle of electromagnetic induction. Electromagnetic induction is defined as the process s of generating voltage in a conductor by changing the magnetic field around the conductor. One of the most effective ways of producing electromagnetic induction for energy harvesting is with the help of permanent magnets, a coil and a resonating cantilever beam.\\
Electromagnetic induction provides the advantage of improved reliability and reduced mechanical damping as there would not be any mechanical contact between any parts; also, no separate voltage source is required. However, electromagnetic materials are bulky in size and are complicated to integrate ~\cite{Roundy}.
\subsection{\large Photovoltaic Cells}
A photovoltaic cell is a device that converts light energy into electrical energy.  The form of energy exploited is typically light energy obtained usually from sunlight.  For locations where the availability of light is guaranteed and usage of batteries and other means of power supply are not feasible or expensive, the usage of photovoltaic cells is a convenient solution.  The most popular photovoltaic cells are the silicon-based cells. These are more sensitive to light, are easily available and offer a reasonable price to performance ratio.\\
The added advantage of using energy harvesting photovoltaic devices is that they are usually small. However, they only scavenge energy from the surroundings and , the supply of energy may be interrupted at a period of time since the power obtained from the surroundings cannot be guaranteed all the time.  Also, the average power available is typically low for such energy harvesters \cite{Raghunathan}.

\subsection{\large Radio Frequency Energy Harvesting}
During radio frequency (RF) based EH, transmitted radio waves are received by a device antenna and converted into a stable AC or DC power source to supply a sensor device.  The energy of radio frequency waves decreases with distance from the transmission source, therefore, the source transmission power, antenna gain, and distance between source and receiver are factors that affect how much energy can be harvested \cite{Shaikh}.  RF methods can be compared according to their conversion efficiency of electric field strength to DC energy for which efficiency usually varies between 50-75 $\%$  ~\cite{Shaikh}.
Despite restrictions on its availability for safety, the fact that RF transceivers are used for both communication and receiving power presents their most unique advantage.

\section{Commercial Applications} 
With the efforts of researchers and engineers, there exits a wide range of self-powered devices. These devices achieve energy autonomous operation merely relying on the energy harvested from the environment.  Here is some examples of the variety of such applications:\\
1. Smart buildings.\\
2. Wearable devices.\\
3. Transportation.\\
4. Implantable medical devices.\\
5. IoT Devices.\\
To summarize, self-powered devices have already begun to appear in a variety of application domains, and new companies are constantly innovating and implementing the concept of energy harvesting. Fig.~\ref{uavenergycity} depicts the vision of cities with UAV-assisted energy harvesting.

\vspace{1cm}
\begin{figure}[h]
    \centering
    \includegraphics[scale = 1.3]{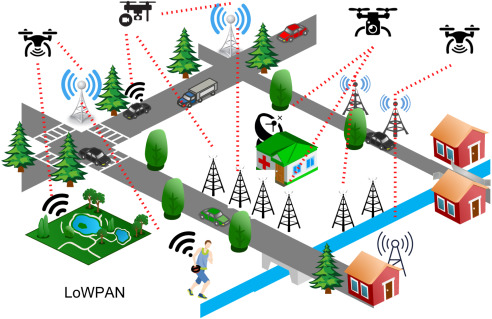}
    \caption{UAV assisted energy harvesting enabled city}
    \label{uavenergycity}
\end{figure}
\newpage

\section{Energy Harvesting in Wireless Communications}
Energy harvesting models are critical in the design of energy scheduling and the evaluation of the performance of energy harvesting wireless communications. The models are primarily divided into deterministic models, stochastic models, and other special models based on the availability of non-causal knowledge about energy arrivals at the transmitters.

\subsection{Deterministic Models}
In deterministic models, full knowledge of energy arrival instants and amounts is assumed to be known in advance by the transmitters.\ Deterministic models are useful for characterizing optimal energy scheduling strategies, providing insights into designing some sub-optimal approaches that only require causal energy state information, and benchmarking the fundamental performance limits of energy harvesting systems by assuming that non-causal energy state information is acquired perfectly. Nonetheless, the success of the energy management utilizing this model heavily depends upon on accurate energy profile prediction over a somewhat long time horizon, and modeling mismatch often occurs when the prediction interval becomes enlarged. Hence, the deterministic models are suitable for the applications with the energy sources whose power intensities are predictable or vary slowly\cite{Reddy}.
\subsection{Stochastic Models}
Stochastic energy harvesting models consider energy renewal processes to be random processes. One significant advantage of such models is that non-causal knowledge of energy state information is not required, making them suitable for applications where the energy state information is unpredictable. However, the problem of modeling mismatch  always occurs because it is difficult to fully understand the stochastic behavior of ambient energy sources. In \cite{michel} the energy generation process is described via Bernoulli models with a fixed harvesting rate under the assumption that energy harvested in each time slot is identically and independently distributed.  Other models applied uniform process such as, Poisson process~\cite{mao}, and exponential process~\cite{lei}. While these models are simple, they are inadequate to capture the temporal properties of the harvested energy for most energy sources.\\
In addition, an appropriate choice of the underlying parameters in stochastic models such as the transition probabilities of states and the probabilities of energy arrival amounts at given states is another crucial issue. In real applications, this should be closely related to real empirical energy harvesting data measured by the energy harvester of each communication device, and the energy harvesting capability is typically device-specific.

\subsection{RF Based Models }
Apart from the natural renewable energy sources, a new emerging solution is to collect energy from RF signals which are artificially generated by other external communication devices. This model is commonly used for ambient RF sources, where the RF transmitters are not intended for energy transfer, the model becomes more complicated because the ambient RF transmitters work periodically and their transmit power varies significantly from $10^{6}$ W for TV towers to 0.1 W to WiFi devices. ~\cite{Flint} characterized the average RF energy harvesting rate at devices powered by ambient RF sources. Although the RF energy sources could be deterministic or random, the amount of the harvested energy from RF signals largely depends on two crucial factors: transmit power of dedicated or ambient transmitters and the channels (including path loss, shadowing and small-scale fading) from the transmitters to the harvesting receivers. These two factors make the RF energy sources very different from other “natural” energy sources and introduce a performance trade-off between information and energy transfer in wireless networks.

%============================End of chapter 2=======================
%============================ Start of chapter 3====================
\chapter{Modeling and Problem Formulation}
In this section. The network model, which depicts the overall network equations and parameters will be discussed first.  Following that, we will define the problem we aim to solve. It is worth mentioning that, although the project extends the work of~\cite{Yu2021} from modeling, implementation, performance analysis, discussions and evaluation against some state-of-art,  a reader may expect to see an overlapping between their model,  formulas,  and defined parameters and the ones presented in this chapter. \\First of all, we provide the notations that we are going to be used throughout this project in Table 3.1.
\begin{table}[H]
    \centering
    \begin{large}
        \caption{ SYSTEM MODEL NOTATIONS.}
    \begin{tabular}{|c|c|}
    \hline
         \textbf{Notation}  & \textbf{Meaning}                   \\ \hline 
         $J$ & \hspace{65px}  Number of IoT Devices \hspace{65px}                         \\ \hline 
         $Lj$ & Data in Queue Waiting to be Uploaded          \\ \hline
         $\bigtriangleup t $& Update interval                    \\ \hline
         $\lambda j(t)$   & Data Generation Rate of Device j    \\ \hline
         $Lmax$           & Storage Capacity of the Buffer        \\ \hline
         $Q$ & The Transmission Data Size                         \\ \hline
         $Qj(t)$ & The Data to be Transferred                   \\  \hline
          $q_{j}^{u}(t)$ & Data Upload Priority of the $jth$ Device \\ \hline
        $v_{o}$ & The Mean Induced Velocity Under Hovering\\\hline

    \end{tabular}
    \end{large}
    
        \label{Table:systemn}
\end{table}

\begin{table}[H]
\centering
\begin{large}
\vspace{50px}
\begin{tabular}{|c|c|}
\hline
         $v(t)$  & UAV Flight Speed at Time $t$                   \\ \hline
         $\theta(t)$ & The Yaw Angle                             \\ \hline
         $P_{0}$ &  The Blade Power While Hovering               \\ \hline
               
         $U_{tip}$ & The Tip Speed of Rotor Blade             \\ \hline
         $P_{i}$ &  Induced Power Under the Hover Condition\\\hline
          $d_{0}$ & Fuselage Drag Ratio           \\ \hline
          $\rho $ & Air Density             \\\hline
          $S$ & Rotor Solidity   \\\hline
          $A$ & Rotor Disc Area \\\hline
          $V_{ME}$ & Maximum Endurance Velocity \\\hline
          $D_{dc}$ &  Data Collection Diameter \\ \hline
          $D_{eh}$ & Energy Harvesting Diameter \\\hline
          $P_{d}$ & Downlink Transmitted Power of UAV \\\hline
          $Fc$ & Carrier Frequency  \\\hline
          $C$ & Speed of Light \\\hline
          $d_{0}$ & Reference Distance \\\hline
          $d_{j}^{-\alpha}$ & Propagation Distance Between UAV and IoT Device J\\\hline
          $\alpha$ & Path loss Exponent \\\hline
          $\theta_{j}(t)$ & The Elevation Angle of UAV and IoT Device in Degree\\\hline
          $d_{j}(t)$ & Distance Between UAV and the IoT Device $J$\\\hline
          $h_{j}(t)$ & Downlink Channel Power Gain\\\hline
          $g_{j}(t)$ & Uplink Channel Power Gain\\\hline
          $P_{Limit}$ & The Maximum Output Power \\\hline
          $\sigma^{2}$ & Channel Noise Power\\\hline
          $W$ & Wireless Communication Bandwidth \\\hline
          
         \hline 
    \end{tabular}
\end{large}
        \label{Table:1}
\end{table}
\newpage
\section{Network Model} 
As shown in Fig.~\ref{fig: UAVenv}, we consider a wireless network with a single UAV working as a BS to provide services to IoT devices, Let $j$ denote the IoT devices where $j \in \{1,2,...,J\} $. The devices are located randomly on the ground where $[x_{j}(t) , y_{j}(t)]$ denotes the initial position of the device $j$. 
%================
\begin{figure}[H]
\centering 
\includegraphics[scale = 0.3]{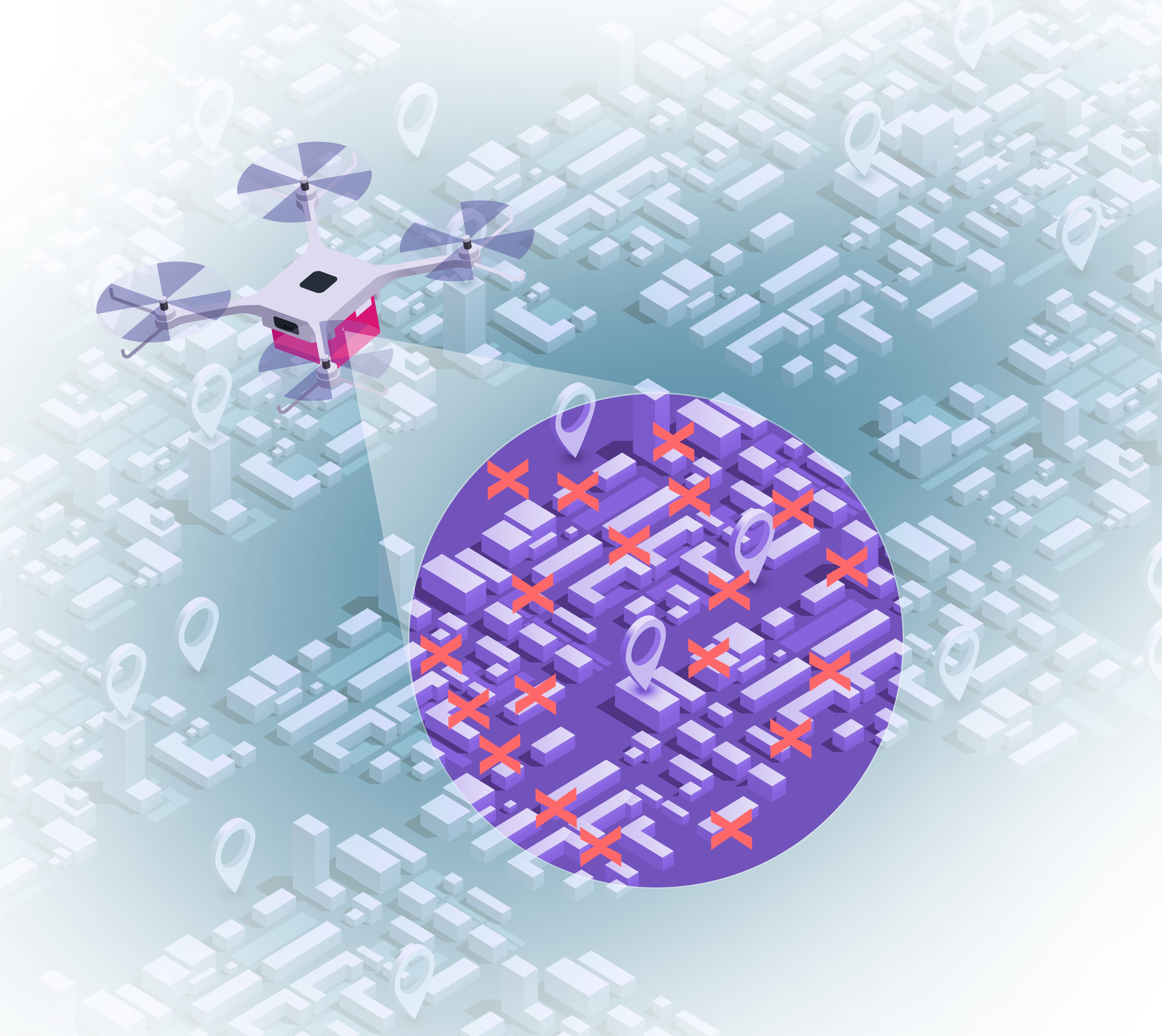} 
\caption{UAV Communication network environment}
\label{fig: UAVenv}
\end{figure}

\section{Data Generation Model}
We assume that the IoT devices monitor their environment and obtain information about their observations, which then are stored in the devices data buffer in real time. Let $L_{j}(t)$ denotes data in a queue waiting to be uploaded at time  $t$, 0 $\le$ $t$ $\le$ T. At time $t+\bigtriangleup t$, $\bigtriangleup t$ is the time to update this data, the buffer data   updates according to 
\begin{equation}
 Lj(t+\bigtriangleup t ) = L_{j}(t) +\lambda_{j}(t)\bigtriangleup t,
 \label{eq1}
\end{equation} 
where   $\lambda_{j}(t)$ is the data generation rate of device $j$ at time $t$. We assume that generation rate of the data  $\lambda_{j}(t)$ obeys to the Poisson distribution, and the parameters of the  Poisson distribution is different for each device. The maximum of $ L_{j}(t)$ is typically constrained by hardware limitation and it is assumed to be bounded by $[0,L_{MAX}]$, where $L_{MAX}$ is the storage capacity of the data buffer and it is assumed to be the same for all devices.\\
When the buffer is full with data, the older data may be overwritten by new data, or recently gathered data may be dropped. As a result, it is of great importance for IoT devices to upload the data collected in their buffers in time and the data to be transmitted size in bits corresponding to $L_{MAX}$ is $Q$.  Therefore the data to be transferred at $t$ is
\begin{equation}
 Qj(t) = \frac{l_{j}(t)}{L_{MAX}} Q.
 \label{eq2}
\end{equation} 
Since the data generation rate vary from device to device,  to manage their priorities efficiently we introduce, similar to \cite{Yu2021},   $q_{j}^{u}(t)$ to represent the data upload priority which is defined as follows.
\begin{equation}
 q_{j}^{u}(t) = \lambda _{j}(t) \frac{l_{j}(t)}{L_{MAX}}
 \label{eq3}
\end{equation}
As we note,  the data transmission priority not only relies on the ratio
of gathered data to the storage capacity, but is also affected
by data generation rate.

\section{Users Mobility Model}
To put our model to the test in a more realistic settings, we added mobility to a randomly selected set of devices. The model considered is 2-D random-walk square cell~\cite{shenoy}, which allows devices a move uniformly randomly over four pre-specified locations.  In contrast to~\cite{shenoy}, however, the number of available locations for each mobile device in our model are made  significantly large to ensemble a continuous spatial movement.

\section{UAV Model}
{
Since the UAV is energy limited, we assume that each flying mission lasts for a  specified period. The UAV adopts  fly-hover-communicate protocol, where  performs data collection and energy transfer processes when hovering at a corresponding location only.\\
We assume   that the UAV operates in full-duplex mode where it transmits energy to IoT devices in downlink and collects data from IoT  devices in uplink simultaneously.\\
\subsection{Mobility}
The UAV  is assumed to fly at a fixed altitude $H >0$ and its horizontal location at time $t$ is denoted as $[x_{u}(t),y_{u}(t)]$. The UAV determines its next action in real time and updates the position accordingly. The flight control of the UAV is described by flight speed $v(t)$ and yaw Angle $\theta(t)$, where $v(t)$ is limited by the maximum flying velocity $Vmax$ and $\theta(t)$ $\in $ $ [-\pi , \pi]$.
}
\subsection{Propulsion Power Consumption}
{
The UAV propulsion power consumption while flying with speed $v$ can be calculated as follows 
\begin{equation}
 P(V)= P_{0}\left(1+\frac{3V^{2}}{U_{tip}^{2}}\right)+P_{i}\left(\sqrt{1+\frac{V^{4}}{4v_{o}^{2}}}-\frac{V^{2}}{2v_{o}^{2}}\right)^{1/2}+\frac{1}{2} d_{0} \rho_{s} A V^{2}
 \label{eq4}
\end{equation}\\
The propulsion power consumption of the  \acrshort{uav} includes blade profile, induced power and parasite power, corresponding to the three parts of the above formula. $P_{0}$ is blade profile power in hovering and  $U_{tip}$ is the tip speed of rotor blade. $P_{i}$ and $v_{o}$ denote induced power and the mean rotor induced velocity under the hover condition, respectively. As for parasite power, $d_{0}$, $\rho $, $S$, and $A$ respectively denote the fuselage drag ratio, air density, rotor solidity, and rotor disc area. The variation trend of propulsion power consumption versus speed is shown in Fig.~\ref{fig:UAVvsSpeed}.
\begin{figure}
    \centering
    \includegraphics{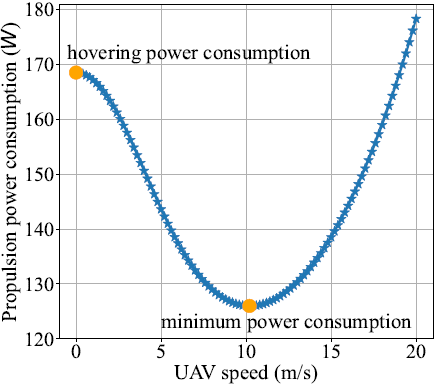}
    \caption{\large Propulsion power consumption versus speed (V) }
    \label{fig:UAVvsSpeed}
\end{figure}\\
As we can note, the power consumption decreases first and then increases with the increase in acceleration and the hovering power consumption $\large P_{hov} = P_{0} +  P_{i}$ can be calculated by setting $V = 0$.\\
\section{Channel Model}
We denote the downlink channel power gain and uplink channel power gain of wireless communication link between \acrshort{uav} and the  $j_{th}$  device as $h_{j}(t)$ and $g_{j}(t)$, respectively. The practical air-to-ground channel model that combined with \acrfull{los} link and \acrfull{nlos} link is considered.  The mathematical description of corresponding path loss is given as follows
\begin{equation}
 L_{j}(t) = \left\{\begin{matrix}
\gamma_{0}~d_{j}^{-\alpha} & , \acrshort{los}~link \\ \mu^{\acrshort{nlos}}~\gamma_{0} ~d_{j}^{-\alpha}
 & ,NLoS~ link 
\end{matrix}\right.
 \label{eq5}
\end{equation}\\
where $\Large \gamma_{0} = \left(\frac{4\pi f_{c}}{c}\right)^{-2}$ represents channel power gain at the reference distance of d0 = 1 m, with $F_{c}$ denoting the carrier frequency and $c$ denoting the speed of light. $d_{j}^{-\alpha}$ is the propagation distance between UAV and IoT device $j$ where $\alpha$ stands for the path loss exponent $\mu^{\acrshort{nlos}}$ is the additional attenuation coefficients of NLoS link. As for IoT device $j$, the \acrshort{los} probability at time $t$ can be expressed as
\begin{equation}
 P_{j}^{Los}(\theta_{j}(t)) = \frac{1}{1+a *exp(-b(\theta_{j}(t)-a))}
 \label{eq6}
\end{equation}
The \acrshort{los} probability of channel condition depends largely on the propagation environment. a and b are constant values that depend on the carrier frequency and the type of environment. It also influenced by the relative location of the communicating parties. $\theta_{j}(t)$ is the elevation angle of UAV and IoT device in degree. It is given as $\theta_{j}(t) = \frac{180}{\pi} \sin^{-1} \left(\frac{H}{d_{j}(t)}\right)$. $d_{j}(t)= \sqrt{(y_{u}(t)-y_{j})^{2}+H^{2}+(X_{u}(t)-x_{j}})^{2}$ is the distance between UAV and the IoT device $j$ The probability of the NLoS component then can be given by $P_{j}^{Nlos}(\theta_{j}(t)) = 1- P_{j}^{los}(\theta_{j}(t))$ . We assume the uplink and downlink channels are approximately equal. As a result, the channel power gain between \acrshort{uav} and \acrshort{iott} device $j $ is given as $h_{j}$
\begin{equation}
 h_{j}(t)  \simeq g_{j}(t)
              =(P_{j}^{Los}(\theta_{j}(t))+\mu^{Nlos}P_{j}^{NLos}(\theta_{j}(t)))\gamma_{0}d_{j}(t^{-\alpha})
 \label{eq7}
\end{equation}
\section{Energy Harvesting Model}
We assume that the range of UAV’s data collection and energy transfer is limited. The UAV only charges and collects data from IoT devices that fall within the coverage. This assumption is reasonable since communication is inefficient when \acrshort{iott} devices are too far away from the UAV. We denote $D_{dc}$ and $D-{eh}$ to represent the maximum coverage radius of data collection and energy transfer respectively.  At each moment, UAV chooses an IoT device as the target device for data collection.  Once the target device falls within $D_{dc}$, UAV will hover at the corresponding location to receive information and transmit energy to other devices within $D_{eh}$ at the same time until the target device completes its data upload.} The received power at a device $j$ located within $D_{eh}$  can be  described 
\begin{equation}
 P_{j}^{r}(t)=\left | h_{j}(t) \right |^{2} P_{d}, %\forall \triangle \le D_{eh},
 \label{eq8}
\end{equation}
where we denote $P_{d}$ to represent the downlink transmit power of the UAV. 
In our work , we apply the non-linear \acrshort{eh} model ~\cite{Ng}.  Different from linear model, non-linear \acrshort{eh} model considers the saturation limitation of the circuits and is more practical. The harvested energy is described by
\begin{equation}
 P_{j}^{h}(t)=\frac{P_{limit} e^{cd}-P_{limit} e^{c(P_{j}^{r}(t)-d)}}{e^{cd}(1+e^{-c(P_{j}^{r}(t)-d)})}
 \label{eq9}
\end{equation}
Where $P_{limit}$ is the maximum output power, $c$ and $d$ are constants that depend on related circuit characteristics of the \acrshort{eh} system.

\newpage
\section{Problem Formulation}
In this project, our aim is to optimize the maximum sum data rate and total harvested energy of \acrshort{iott} devices, and minimize energy consumption of \acrshort{uav} at the same time. The \acrshort{uav} is required to observe the \acrshort{iott} environment and implement a real-time path planning, the decision of \acrshort{uav} flying trajectory and the choosing of hovering location should consider the \acrfull{qoss} of the devices and the \acrshort{uav} and the energy consumption of the \acrshort{uav}, However , avoiding data overflow of all \acrshort{iott} devices is of great importance, Therefore , The UAV should successively visit the devices according to their requiring priority. For example, the \acrshort{iott} device $j= arg_{j} ~max$ $q_{j}^{u}(t)$ will be chosen as the target device of \acrshort{uav} at $t$. When the UAV flies to the range $d_{j}(t) \leq D_{dc}$ it starts hovering at the corresponding location to start the data collection process in the uplink and transmitting energy in the downlink, let $k$, $0 < k \leq K$ represent the $k^{th}$ hovering of the UAV in a mission. Where $ K \geq 0$ denotes the total number of times that UAV hovers to communicate with IoT devices. We denote the corresponding communication device of the $k_{th}$ hovering as $j_{k}$. Then the transmission data rate at the $k_{th}$ hovering is given as
\begin{equation}
 R^{k} = W log_{2}\left(1+\frac{P_{u}|g_{j}^{k}(t)|^{2}}{\sigma^{2}} \right )
\label{eq12}
\end{equation}
where $W$ is the wireless communication bandwidth and $\sigma^{2}$ is the channel noise power at UAV. To upload all of the gathered data to UAV the hovering time can be calculated by
\begin{equation}
 t^{k}=\frac{Q_{j}^{k}(t)}{R^{k}}
\label{eq13}
\end{equation}
At the same time while the UAV is gathering data in the uplink, in the downlink it keeps transmitting energy to the devices within its $D_{eh}$ except the device the data is gathered from, the harvested power at device
$j$ is given by
\begin{equation}
 E_{j} = P_{j}^{h}(t)t^{k} ,   \forall \triangle d_{j}(t) \le D_{eh}  , j \neq j^{k} 
\label{eq14}
\end{equation}
and the total harvested energy at the $k^{th}$ hovering is given as
\begin{equation}
 E^{k} =\sum_{\forall \triangle d_{j}(t) \le D_{eh} , , j \neq j^{k}} E_{j}
\label{eq15}
\end{equation}
The sum data rate and the total harvested energy of all the hovering stages in a mission are given as following.
\begin{equation}
 R_{sum} = \sum_{k=0}^{k} R^{k}
\label{eq16}
\end{equation}
\begin{equation}
 E_{total}^{h} =\sum_{k=0}^{k} E^{k}
\label{eq17}
\end{equation}
And the total energy consumption for UAV’s flying and hovering in the task duration is given as
\begin{equation}
 E_{total}^{c} = \int_{0}^{T}P(v(t))dt.
\label{eq18}
\end{equation}
As we noted , that energy consumption also includes communication energy. Since we assume the downlink transmit power $p_{d}$ a constant, this component is not included in the optimization objective the \acrshort{moo} problem can be formulated as
\begin{equation}
 P1:  max_{v(t),\theta(t)}(R_{sum},E_{total}^{h},-E_{total}^{c})
\label{eq19}
\end{equation}
\begin{equation}
 st: v(t)\in [0,v_{max}]
\label{eq20}
\end{equation}
\begin{equation}
 \theta (t) \in [-\pi,\pi]
\label{eq21}
\end{equation}

For the sum data rate, its optimization depends on the number of devices that requiring to upload data over the UAV mission period that is, the total number of hovering for the UAV. There is no need to say that to optimize the $R_{sum}$
The UAV should fly at a higher speed to visit more IoT devices and its hovering location should be as close to the target device to data rate. From This point of view, hovering over the target is the best choice. However, there is a chance that the target has some sort of mobility, since it is not practical to assume that all of the targets are always at a fixed location, therefore, the hovering location should consider being close to the target and the \acrshort{qoss} for the devices. \\
As for the optimization of the total energy harvested, we hope that more devices falls within the range whitin each UAV hover.  In addition, the smaller the distance between UAV and energy collecting devices the better.  This may conflict with the UAV hovering location for the data collection target, thus conflicting with the optimization of the sum data rate.\\
As for the objective of UAV’s energy consumption, it is clearly that the velocity corresponding to the minimum energy consumption can achieve this minimization.  However, this velocity doesn't guaranty that it would be fast enough to collect data from the target devices without them facing the overload problem, let alone maximizing the number of visited devices.\\
As we can see, these three objectives are in conflict with each other partly.  Since the devices are randomly distributed for each instance and their data generation are dynamic, it is substantially complex and may impose considerable computational cost to find out an optimal hovering location and make flying decision.  Furthermore, traditional model-based methods like dynamic programming method are unable to fix this problem.  Recently, DRL has shown excellent ability of solving complex problems and is regarded as one of the core technologies of artificial intelligence.  As the integration of deep learning and RL, it owns the strong understanding ability and decision-making ability and thus can realize end-to-end learning. It has shown great potential in solving sophisticated network optimizations.  DDPG has been proved that can learn effective polices in continuous action spaces using low dimensional observations.  It is suitable for our proposed UAV’s flight decision problem where flying speed and yaw angle are chosen in continuous interval.  Since the reward of original DDPG algorithm is scalar, we extend it to multidimensional reward for the Multi-objective optimization problem.

%==================================== End of chapter 3====================
%======================Start of Chapter 4 ========================
\chapter{DDPG for Green Mobile IoT  Networks }%{Deterministic Policy Gradient}
Policy gradient algorithms are widely used in reinforcement learning problems with continuous action spaces . The basic idea is to represent a policy ($\pi$)  by a parametric probability distribution $\pi_{\theta}(a|s;\theta)$ that selects actions ($a$) in state ($s$) according to a parameter vector ($\theta$).  Policy gradient algorithms typically proceed by sampling this policy and adjusting its parameters in the direction of greater cumulative reward ($r$). \\
When studying reinforcement learning and control problems the model is usually based on \acrfull{mdp} which comprises: a state space ($S$), an action space ($A$), an initial state distribution with density $p_{1}(s1)$, the policy is used to select actions in the MDP and the agent The agent uses this policy to interact with the MDP to give a state, next action and reward.  The agent’s goal is to obtain a policy which maximises the reward from the start state.
\section{Actor-Critic Algorithms}
The actor-critic is a widely used architecture based on the policy gradient theorem~\cite{actorcritic}.  It consists of two eponymous components.  An actor which adjusts the parameter $\theta$ of the stochastic policy $\pi_{\theta}(s)$ by gradient ascent of Equation \ref{grascend}.  Instead of the unknown action-value function $Q^{\pi}(s,a)$ in Equation \ref{grascend} an action-value function $Q^{w}(s,a)$  is used, with parameter vector $w$.
\begin{equation}
  \nabla_{\theta}J(\pi_{\theta})=\int \rho^{\pi}(s)\int \nabla_{\theta}\pi_{\theta}(a|s)Q^{\pi}(s,a).dads  
 \label{grascend}
\end{equation}
 A critic estimates the action-value function $Q^{w}(s,a) $ $ \simeq$  $Q^{\pi}(s,a)$ using an appropriate policy evaluation algorithm.\\
So far, the actor produces a stochastic policy assigning probabilities to each discrete action or necessitating sampling in some distribution for continuous actions .  However , There are two drawbacks:
\begin{enumerate}
  \item the value of an action estimated by the critic must have been produced recently by the actor, otherwise the bias would increase dramatically which  
  \item Because of the randomness of the policy, the returns may vary considerably between two episodes generated by the same optimal policy.  This induces a lot of variance in the policy gradient, which explains why policy gradient methods have a worse sample complexity than value-based methods: they need more samples to get rid of this variance.
\end{enumerate}
In the next section, we will see the state-of-the-art method DDPG (Deep Deterministic Policy Gradient), which tries to combine the advantages of policy gradient methods (actor-critic, continuous or highly dimensional outputs, stability) with those of value-based methods.
\section{ DDPG Algorithm }
The basic principle behind DDPG is to extend the Deterministic Policy Gradient approach to work with non-linear function approximators~\cite{DDPGP} by simply using a memory where  the agent's experiences at each time step is stored in a data set called the replay memory and utilizing target networks to stabilize the learning process.  However, one issue remains, as the policy is deterministic, it can very quickly produce always the same actions, missing perhaps more rewarding options.  Some environments are naturally noisy, enforcing exploration by itself, but this cannot be assumed in the general case.  The solution retained in DDPG is an additive noise added to the deterministic action to explore the environment called exploration noise. 
% \vspace{1cm}
\begin{figure}[h]
    \centering
    \includegraphics [width=0.6\textwidth]{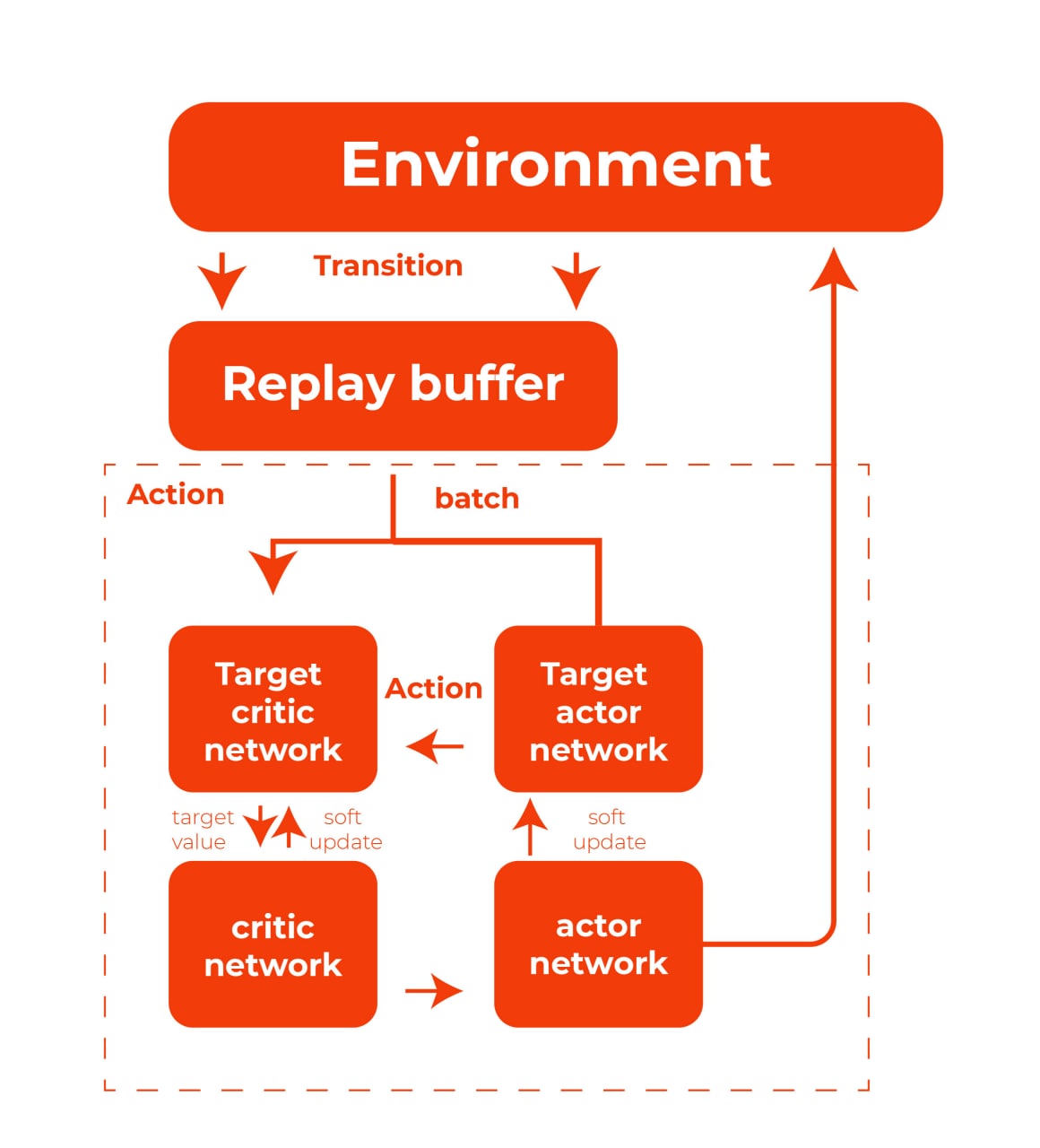}
    \caption{\large Basic Framework of DDPG algorithm}
    \label{fig:ddpgfw}
\end{figure}
\\
It is worth to mention that DDPG algorithm is off-policy algorithm where the samples used to train the actor come from the replay buffer and there is no need to balance the probabilities of the behavior and target policies, In other words, the importance sampling weight can safely be set to 1 for DDPG.\\
DDPG has rapidly become the state-of-the-art model-free method for continuous action spaces due to it capability  to learn efficient policies on most continuous problems, either pixel-based or using individual state variable.

\section{DDPG Algorithm for  UAV-Assisted Energy Transfer and Data Collection}
In this section, we first build the environmental model based on the model of \cite{Yu2021} to map the system model to the interaction environment of MDP.  And then propose an UAV-assisted data collection and energy transfer algorithm for the proposed problem.
\subsection{Environmental Model}
As we mentioned, the agent agent depends on the interaction with the environment to adjust its behavior and learn optimal policies.  Therefore, it is of great importance to cast the optimization problem into the MDP in a right way.  In this section we give detail description of the design of state space, action space and reward in our model.
\subsubsection{\large State Space}
Collecting the real-time service requirements of all the IoT devices relies on frequent information exchange between the UAV and IoT devices.  It will occupy a large amount of wireless resources and cause delay, greatly reducing the efficiency of the system.  To be more practical, we assume that the UAV can only observe its own state and partial network information.  To be specific, UAV can observe its own location, the cumulative number of flights out of the restricted area, the location of the target device and the number of devices with data loss.  And then the state space is defined symbolically as
\begin{equation}
     S  \triangleq {s_{t}} ={[d_{J}^{x}(t), d_{J}^{y}(t), x_{u}(t),y_{u}(t),N_{f}(t),N_{d}(t)]}
\end{equation}
where $[d_{J}^{x}(t),d_{J}^{y}(t)]$ is the distance between the target device and the UAV under the cartesian coordinates.  Once the UAV has finished the data collection of the target device, a new one will be selected according to the status of the system at the time. This element helps to guide the UAV to get the target devices into its data collection coverage.   $N_{f}(t)$ records the cumulative number of times that the UAV has continuously exceeded the restricted area by the time $t$.  Combining with UAV’s absolute position $[x_{u}(t) , y_{u}(t)]$ it helps to keep the UAV from flying out of the designated area that causes unnecessary waste of resources.  And the number of devices with data loss $N_{d}(t)$ will drive the UAV to service the high-demand devices timely.  In practical scenarios, the global network information is incapable to obtain and the real-time knowledge about each device is unknowable at the UAV. Besides, most of the information is not necessary for decision-making.  In our setting, we extract a small amount of necessary information to represent the state of the environment.  These elements of state space will enable the UAV to have a good overall perception of the environment.  Furthermore, it overcomes the lack of network information which is common problem that exists in massive uncertain IoT system.
\subsubsection{\large Action Space}
Observing the state, the UAV makes action decision in real time.  The action space is defined as
\begin{equation}
     A \triangleq a_{t} =[{v(t)cos(\theta(t)), v(t)sin(\theta(t))}]
\end{equation}
We use $[cos(\theta(t)), sin(\theta(t))]$ to represent the yaw angle and then the network will learn a normalized two-dimensional vector.  The flying speed $v(t)$ and the yaw angle $\theta(t)$ are assumed to be continuous value in the interval $[0,v_{max}$ and $[-\pi,\pi]$ respectively.  It enlarges the control freedom of the UAV as well as improves the efficiency of the control scheme comparing to discrete action space.
\subsubsection{\large Reward}
Since the environment is partially observed, the UAV depends on the reward to evaluate its decision, infer the distribution of states and learn and know the environment. Besides, the agent relies on the well-designed reward function to learn effective control policy for the proposed MOO problem.  According to our optimization problem, the reward is designed as a 4-dimensional vector.
\begin{equation}
     R \triangleq {r_{t}} ={[r_{dc}(t),r_{eh}(t),r_{ec}(t),r_{aux}(t)]}
\end{equation}
where $r_{dc}(t)$, $r_{eh}(t)$, $r_{eh}(t)$ correspond to the three optimization objectives: maximization of sum data rate, maximization of total harvested energy and minimization of UAV’s energy consumption.  They are designed as following.
\begin{equation}
     r_{dc}(t)=\left\{\begin{matrix}
  w_{dc} \times R^{k},& k^{t} hovering \\ 0,
 & otherwise
\end{matrix}\right.
\end{equation}
\begin{equation}
 r_{eh}(t)=\left\{\begin{matrix}
w_{eh}\times (E^{k} + \sum_{0}^{j}I_{d_{j}(t)}),& k^{t} hovering \\ 0,
 & otherwise
\end{matrix}\right.
\end{equation}
\begin{equation}
 r_{ec}(t)=\left\{\begin{matrix}
w_{ec}\times -P_{hov},&  hovering \\ w_{ec}\times-P_{v(t)},
 & otherwise
\end{matrix}\right.
\end{equation}
Where $\{w_{dc}, w_{eh}, w_{ec}\} $ are the  weights of the optimization problems corresponding to data collection,energy harvesting and energy consumption respectively. These weights are designed to emphasize on the importance of an optimization objective.\\
Once the target device falls within the data collection coverage radius of UAV, the UAV will hover to process data collection and energy transfer.  Otherwise the UAV is in flying stage.  We give more rewards to the agent for its higher data rate, more harvested energy at more IoT devices in hovering, and punish it for its higher energy consumption at both flying and hovering stages.  $w_{eh}$, $w_{dc}$ and $w_{ec}$ are priority weights associated with each attribute.  In addition, there is an auxiliary reward $r_{aux}(t)$ that given as
\begin{equation}
 r_{aux}(t) = -d_{j}^{x}(t)-d_{j}^{y}(t)-N_{f}(t)-N_{d}(t)
\end{equation}
It can be seen that $r_{aux}$ includes the distance between UAV and the target device. It will be small if the UAV is far away from the target device, which helps the UAV recognize the location of the target device so as to get close to it.  Besides, if the UAV tries flying out of the restricted area or leads to IoT devices’ data overflow due to the failure of timely data collection, it will get negative reward. We inflict punishment on UAV’s bad flight decisions to drive the UAV to learn to finish the basic tasks no matters the preferences of the optimization objectives. The corresponding weight $w_{aux}$ is set as 1 all the time.
\vspace{5cm}
\section{MODDPG Algorithm }
The basic frame work of the DDPG algorithm is presented in Fig.~\ref{fig:ddpgfw}. Based on the DDPG architecture.  we maintain an actor network $\mu(s|\theta^{u})$ to specify the main policy that builds a mapping from states to actions and a critic network $Q(s|\theta^{u})$ to estimate the action value.  $\theta^{\mu}$ and $\theta^{Q}$ are parameters of two networks.  The weights of both the actor network and critic network are initialized from a truncated normal distribution centered on 0 with standard deviation $\sqrt{2/i}$, where $i$ is the number of input units in the weight tensor.  The biases are all initialized as 0.001. Besides, target network is applied to the actor-critic architecture to calculate the target values. Specifically, a target actor network and a target critic network are created by copying the parameters of main network in the initialization phase.\\
While updating the network parameters, a random mini batch of experience tuples are sampled uniformly from replay memory.  Different from the original DDPG, which is single objective MDP with scalar reward signal, the reward in the experience tuples is a vector.  Since the value of the action depends on the preferences among competing objectives, we use the linear weighting method to calculate the weighted sum of elements of the reward vector with the given weights, which is given as $r=rw^{T}$, where $w=[w_{dc},w_{eh},w_{ec},w_{aux}]$ Then the reward vector is transformed into scalar form.\\
\vspace{-11cm}
\begin{algorithm}[t]
\begin{large}
\caption{MODDPG algorithm for UAV-assisted energy Harvesting and data collection  }\label{alg:cap}
 \hspace*{\algorithmicindent} \textbf{Input} a weight vector $w =$ $[w_{dc}, w_{eh}, w_{ec}, w_{aux}]$.\\
 1 : Initialize main network and target network;\\
 2: Initialize replay memory $\beta$, Initialize $\sigma^{2} =2.0$, $\epsilon= 0.9999$ for action exploration;\\
 3: for episode :$=$ 1,...,M \textbf{do}\\
 4: \hspace{6pt}    \textbf{for} step t: $=$ 1,...,T \textbf{do}\\
 5: \hspace{9pt} Update the environment status and observe the current state $s_{t}$\\
 6: \hspace{19pt} Select action according to $a$ $\sim$ $N(\mu(s_{t}|\theta^{\mu},\epsilon \sigma^{2})$\\
 7: \hspace{28pt} Execute action $a_{t}$ and limit UAV in designated area, observe reward $r_{t}$, transit to the next \\\hspace{35pt} state $s_{t}+1$;\\
 8: \hspace{30pt} Store the experience tuple $(s_{t}, a_{t},r){t}, s_{t+1})$ into replay memory $\beta$;

If{update}
    State Randomly sample a mini-batch transitions from $\beta$ .\\
    State Compute $y_{i}$
    State Update critic network by minimizing the critic loss\\ 
    State Update actor network by maximizing the actor loss\\

\hspace{50pt}\textbf{end if}\\
\hspace{40pt} \textbf{end for}\\
\textbf{end for}
\end{large}
\end{algorithm}\\
\newpage
It should be noted that through this design, the MODDPG algorithm is suitable for MOO problem with arbitrary number of objectives.  And it also supports single objective optimization.\\
To optimize the main critic network, we calculate the difference between target value and Q-function given by the main critic network. Then the main critic network is trained by using the gradient descent method to minimize the loss function, which is defined as the mean square error of the difference.
\begin{equation}
     L(\theta^{Q}) = E [(Q(s_{i},a_{i}|\theta^{Q}-y_{i})^{2}]
\end{equation}
The loss function of actor network is simply obtained by calculating the sum of Q-function for the states.  We use main critic network and pass action computed by main actor network to compute the Q-function.  The loss function of actor network is
\begin{equation}
    L(\theta^{\mu}) = E [(Q(s_{i},\mu(s_{i}|\theta^{\mu})|\theta^{Q})
\end{equation}
The chain rule is applied to update actor network weights by maximizing $L(\theta^{\mu})$.  And the parameters of two target networks will update during the training using “soft” target update.  The complete algorithm is presented in \textbf{Algorithm\ref{alg:cap}}.

%======================== End of Chapter 4 =======================
%================================ Start of Chapter 5============
\chapter{Fast and Green Communications}
In this chapter, we  evaluate our model performance and  compare it with  some other baselines.
\section{Simulation Settings}
In our simulation, for actor network, we used 4 fully connected hidden layers beside the input and output layers, for each of these hidden layers the number of neurons is set to $\{400,300,300,300\}$, the activation function used for all of these hidden layers is ReLU, the output layer neurons has two different activation functions which are sigmoid for velocity and tanh for the angle. For the critic network we used 2 fully connected hidden layers, the number of hidden layers is set to $\{400,300\}$, the activation function used for both layers is ReLU, the output layer of the critic network has no activation function. We set the number of IoT devices, randomly distributed in a square area with the range of 400 m by 400m,  to 100 with the selection of 30 device to be having a mobility.  We set the mission period to 10 minutes. At the beginning of each task, the UAV begins its mission at a random position in the designated area. The UAV is assumed to be flying at an altitude of 10 m and its maximum flying speed $v_{max} = 20m/s$. The radius of UAV’s coverage are set to $D_{dc} = 10m$ and $D_{eh} = 30m$. The transmit power of the UAV and IoT devices are set to $P_{d} = 40 dBm$ and $P_{U} = -20dBm$ respectively. IoT device buffer data is updated every second. Their Poisson process of data accumulation expectations are randomly set from these values $[4,8,15,20]$. The data buffer has a capacity of $C = 5000$ packets. The size of the data to be transmitted is set to $ Q=10Mbits$. Other system and algorithm simulation parameters are listed as follows in Table\ref{simps}.
\begin{table}[H]   
    \centering
    \begin{large}
    % ?\end{Large}
        \caption{ SIMULATION PARAMETERS}\label{simps}  
        \vspace{5px}
    \begin{tabular}{|c|c|}
  
    \hline
          Bandwidth $B$ & 1MHz                 \\ \hline 
          Noise Power $(\sigma_{n}^{2})$ & -90dBm \\\hline
          Reference channel Power Gain $(\gamma_{0})$ & -30dB\\\hline
          Attenuation Coefficients of NLoS Link $(\mu)$ &  0.2\\\hline
          Path Loss Exponent ($\alpha$) & 2.3 \\\hline
          Parameters of LoS Probability(a, b) & 10 , 0.6\\\hline
          Blade Profile Power ($P_{o}$) & 79.86W \\\hline
          Induced Power ($P_{i}$) &  88.63W \\\hline 
          Tip Speed of Rotor Blade $(U_{tip})$ & 120m/s \\\hline
          Mean Rotor Induced Velocity in Hover & 4.03m/s \\\hline
          Fuselage Drag Ratio $d_{0}$ & 0.6\\\hline
          Air Density $\rho$ & 1.225 km/$m^{3}$\\\hline
          Rotor Disc Area $(A)$ & 0.503$m^{2}$ \\\hline
          Maximum DC Output Power $(P_{limit})$ & 9.079 $\mu W$\\\hline
          EH Parameters $(c,d)$ & 47083, 2.9 $\mu W$ \\\hline
          Actor Network & [400,300]\\\hline
          Critic Network & [400,300]\\\hline
          Number of Training Episodes & 1600 \\\hline
          Learning Rate for Actor & $10^{-3}$\\\hline
          Learning Rate for critic & $10^{-3}$\\\hline
         
    \end{tabular}
    \end{large}
    
\end{table}
\newpage

\section{Learning Performance-Training}
Approximately, at about the 15 epochs, the accumulated reward fluctuates at a very low level because in this stage, the UAV is in complete experimenting stage and have a small amount of experience to learn from at this stage. Furthermore, all the optimization problems are expected to be not optimized due to the loss of the network is at the levels near the value of (0). As shown in Fig.~\ref{fig: rewa}. the agent quickly learns to obtain higher expected total rewards as training progresses.  And then the accumulated reward converges steadily at a high level. 
\begin{figure}[h]
    \centering
    \includegraphics[width=0.7\textwidth]{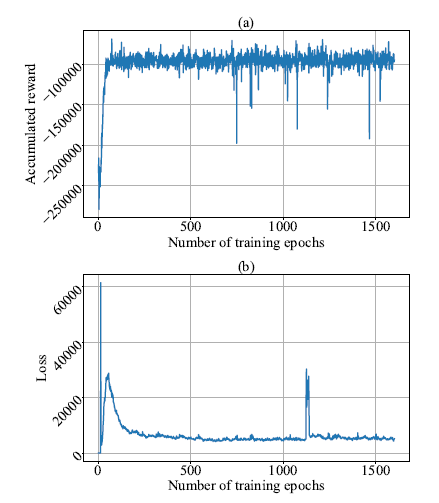}
    \caption{Training curves of the network: (a) Accumulated reward; (b) Loss.}
    \label{fig: rewa}
\end{figure}
\\
When the replay memory is full, the UAV starts to sample its stored experience to train the network.  We can notice an exploration and learning stage before the mark of the 500th episode. During this stage the network loss decreases rapidly after its sharp rise indicating that the UAV started learning process of optimizing the objectives and the sum data rate as well as the total harvested energy increase rapidly while there is a rapid decrease in the UAV energy consumption.
\begin{figure}[H]
    \centering
    \hspace{-0.2cm}
    \subfloat[]{
    \includegraphics[width=0.8\textwidth]{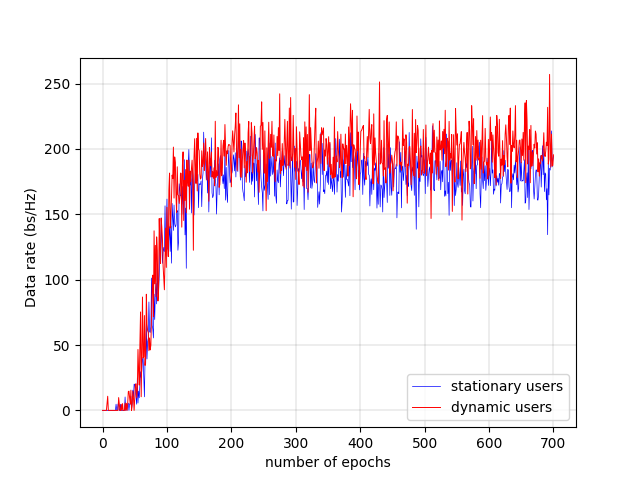}
    }\hspace{0.5cm}
    \end{figure}
    \begin{figure}[H]
    % \ContinuedFloat
    \centering
    \hspace{-0.75cm}
    \subfloat[]{
    \includegraphics[width=0.8\textwidth]{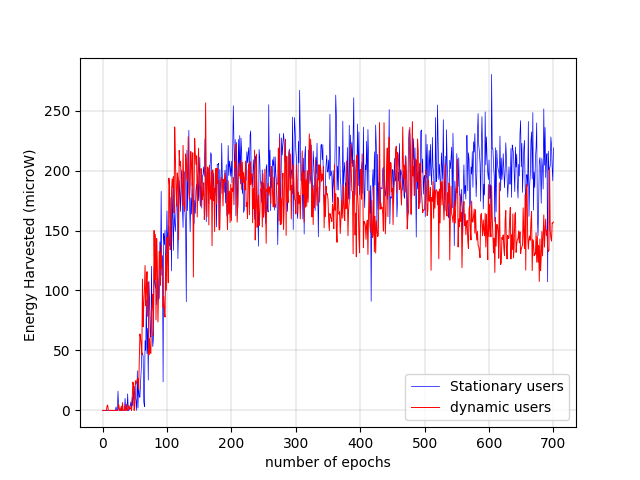}}\hspace{0.5cm}
    \subfloat[]{
    \includegraphics[width=0.8\textwidth]{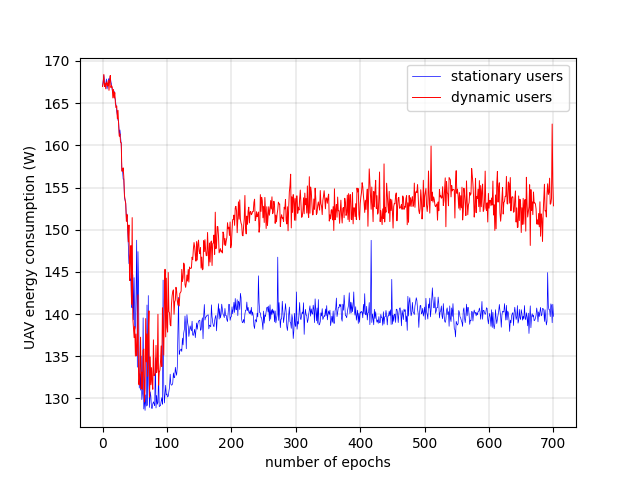}}\hspace{0.5cm}
    \caption{Training curves tracking optimization results: (a) sum data rate; (b) total harvested \centering energy; (c) average energy consumption for dynamic and stationary users}.
    \label{fig:compare}
\end{figure}
 By plotting the data rate over time  trained  in Fig.~\ref{fig:compare} a, we show that the UAV learns to get more targets into $D_{dc}$ to activate the data collection and the average data rate reaches its maximum range quickly for both stationary and dynamic devices. 
As for the harvested energy in Fig.~\ref{fig:compare} b, in the case of the stationary user, the UAV also learns to get more devices in its $D_{eh}$ range, therefore, the average harvested energy keeps improving similar to the case of the data rate.
However, for the dynamic users’ case the average harvested energy declines relative to that for the data rate. The intuition behind this is that, while the agent keeps serving users to meet data rate requirements, the number of users that gets within energy harvesting radius can vary with a chance of having more of these devices get relocated out of this radius and hence reduce the amount of energy harvested. In Fig.~\ref{fig:compare} c For the UAV energy consumption, we can note that the agent learns how to satisfy the energy consumption optimization problem and minimize its average value. However, it is clear the in the case of the dynamic users the UAV energy consumption would be slightly greater that in the case of the dynamic user for the same purpose of tolerating their mobility.  Furthermore, these results shows that optimization problems are learnt with a stable base and that the DDPG technique can produce effective policies to optimize non-convex problems effectively.

\section{Framework Effectiveness- Testing } %$P_{MODDPG}$ Performance vs. Analytical Models
In this section we are going to examine the effectiveness of the DDPG technique.
We train the  algorithm for 1600 episodes to give more importance to either the data rate or energy consumption objective 
by setting the weights of these optimization targets \{ $w_{dc},w_{ec}$\} to be $\{100,1\}$ in the case of having a higher average rate  matters the most which is denoted $P_{SODR}$ and $\{1,100\}$ in the case where the energy consumption  matters the most which is denoted $P_{SOEC}$. In each case we determine the average rate and energy per user and compare to the results presented by~\cite{Zhang2017}, which  solves the maximization of the average data rate  and the minimization of the UAV energy consumption problems using conventional optimization framework. In~\cite{Zhang2017}, similar to $P_{MODDPG}$, the UAV placement is restricted over a 2D plane.  In $P_{MODDPG}$, the theoretical model for the propulsion energy consumption of the UAV is similar to that provided in~\cite{Zhang2017}.  However, different from our work, the authors in~\cite{Zhang2017} assumes the user to be only LoS and static. They  introduced a number of models to optimize rate and energy consumption;  a rate maximization, denoted by $P_{RM}$, energy consummation minimization $P_{EM}$, and  energy efficient $P_{EE}$ models.
 In the $P_{RM}$  model, \cite{Zhang2017} focused on optimizing the UAV trajectory above a target device to ensure the optimal channel condition and hence data rate without taking the consideration of the UAV's flying energy consumption. In the $P_{EM}$, however, the authors solved for the UAV trajectory such that the total energy consumption is minimized. On the other hand, the $P_{EE}$ model jointly optimizes both  average data rate and  the UAV energy consumption with the assumption of circular trajectory centered on the target device. A reader may refer to~ \cite{Zhang2017} for the analytical solution
 provided and more details about these models 
 set  network parameters similar to that   introduced at the begging of the chapter  and present in  Table \ref{evaltable} the average  of rate as well as the consumed energy by the UAV for each model. We also include in the table   the harvested energy determined by the DDPG method. 
 
\begin{table}[H]
\centering
\caption{PERFORMANCE COMPARISON FOR VARIOUS DESIGNS}
\begin{tabular}{|c|c|c|c|}
\hline
      & Average rate (Mbps) & Average power \centering consumption(W) & Harvested energy( $\mu W $)\\\hline
      $P_{RM}$ & 9.56& 585.66 & -\\\hline
     $P_{EM}$ & 2.01 & 102.74 & - \\\hline
     $P_{EE}$ & 7.48 & 118.57 & - \\\hline
      $P_{SODR} $  & 10.7 & 165.7 &  73.07  \\\hline
      $P_{SOEC}$ & 6.48 & 120.76 & 77.98 \\\hline
         
    \end{tabular}

        \label{evaltable}
\end{table}

According to the results we   show in Table \ref{evaltable}, 
we can notice that $P_{SODR}$, achieves a higher date rate than the $P_{RM}$ model with $241.37\%$ reduction in power consumption. When it comes to $P_{SOEC}$, Table \ref{evaltable} illustrates that the results are marginally lower than those of $P_{EM}$ in terms of energy consumption. However, they come  with the benefit of a 222.38$\%$ higher data rate. Finally, $P_{EE}$ when compared to $P_{SOEC}$ for example, we can see that the results are considered interestingly very comparable despite the fact that the trajectory planning is not predetermined and that our method works in more stochastic network compared to that considered in~\cite{Zhang2017} and targets an EH objective which~\cite{Zhang2017} does not consider. 

The results of $P_{MODDPG}$ shows the effectiveness and a degree of freedom compared to the results of the proposed traditional mathematical techniques, even though $P_{MODDPG}$ was in much complex and stochastic environments such as NLoS and dynamic users cases.  $P_{MODDPG}$ learned relatively fast and efficient to solve all the energy management problems we put it in through, despite the fact of them being non-convex.  Overall, AI techniques have the potential to outperform mathematical approaches in many different ways which shows why 6G is called the generation of AI.
%===========================End Of Chapter 4=====================
%============================= Start of chapter 5 ===============
\chapter{Conclusion and Future Work}
In this project, we investigated a multi-objective optimization problem for UAV assisted data collection and energy transfer in mobile IoT networks.  The sum data rate, total harvested energy, and energy consumption were all optimized at the same time.  We investigated a DDPG algorithm to achieve online control of the UAV due to the network's uncertainty and dynamics. The UAV learns to find joint optimization solutions based on weights associated with objectives.  Numerical results validated our algorithm and demonstrated that the model can generate optimized policies. An extension of this work can be investigating the effect of using of multiple UAVs in 3-D environments and the corresponding management challenge which may include interference management and users associations.

%========================== start of References ===================
\bibliographystyle{unsrt}
\bibliography{references}

%======================== End of references======================
\end{Large}
\end{document}